\documentclass{ws-ijmpa}
\usepackage{amssymb,graphicx,amsmath,graphics,subfigure,xfrac}
\usepackage[super,compress]{cite}
\begin{document}

\title{Study of superradiance in the Lambert-W potential barrier}

\author{Luis Puente\footnote{luis.puente@yachaytech.edu.ec}, Carlos Cocha\footnote{carlos.cocha@yachaytech.edu.ec}, Clara Rojas\footnote{crojas@yachaytech.edu.ec}}
\address{School of Physical Sciences and Nanotechnology, Yachay Tech University, 100119 Urcuqu\'i, Ecuador}

\maketitle

\begin{history}
\received{24 April 2019}
\end{history}

\begin{abstract}
We present a new potential barrier that presents the phenomenon of superradiance when the reflection coefficient $R$ is greater than one. We calculated the transmission and reflection coefficients for three different regions. The results are compared with those obtained for the hyperbolic tangent potential barrier and the step potential barrier. We also present the solution of the Klein-Gordon equation with the Lambert-W potential barrier in terms of the Heun Confluent functions.		
\end{abstract}


\ccode{PACS numbers :02.30.Gp, 03.65.Pm, 03.65.Nk}

\section{Introduction}

The Klein-Gordon equation is used to describe spin-0 particles. The scattering of a spin-$0$ particle by a one-dimensional potential barrier is a typical problem that appears in relativistic quantum mechanics and has been the subject of much interest in recent years \cite{schiff:1940,rojas:2005,rojas:2006a, rojas:2006b,rojas:2007,alpdogan:2013,di:2016,rojas:2014a,rojas:2014b,molgado:2018}. Consider an incoming wave from left to right; the common situation is that the wave loses energy because of its interaction with the potential barrier; therefore, the incoming amplitude is greater than the amplitude of the reflected wave \cite{richartz:2009}. The amplitude of the incident wave, $T$, is called the transmission coefficient, and the amplitude of the reflected wave, $R$, is called the reflection coefficient. Another phenomenon, called superradiance\cite{manogue:1988}, where the energy is extracted from the barrier, also appears in relativistic quantum mechanics when the Klein-Gordon equation is applied for an abrupt or smooth potential barrier. In this case, the amplitude of the reflected wave is larger than the amplitude of the incoming wave, which means that more particles are reflected than those that are incident on the potential barrier, or there is a particle--antiparticle pair creation. The superradiance phenomenon has been widely discussed in the literature for the Dirac equation \cite{sauter:1931b,manogue:1988,calogeracos:1999a,calogeracos:1999b,cheng:2009,wagner:2010} and for the Klein-Gordon equation \cite{rojas:2014a,rojas:2014b,molgado:2018}. This phenomenon also appears in astrophysics in the scattering of scalar waves by rotating black holes \cite{richartz:2009,brito:2015}.  

In this article, we study the phenomenon of superradiance, when the reflection coefficient $R$ is greater than one, for the Lambert-W potential barrier. The Lambert-W potential barrier is an asymmetric potential barrier that is solvable for the Schr\"odinger equation for bound states \cite{ishkhanyan:2016e} and scattering states \cite{ishkhanyan:2016d}. The behavior of the reflection $R$ and transmission $T$ coefficients is studied for three different regions of energy: $m<E<V_0-m$, $V_0-m<E<V_0+m$, and $E>V_0+m$. We have observed that for the region $m<E<V_0-m$, $R>1$ and $T<0$; therefore, the phenomenon of superradiance also is observed in the Lambert-W potential barrier. In addition, we compare the reflection $R$ and transmission $T$ coefficients for the Lambert-W potential barrier with those for the hyperbolic tangent potential barrier and the step potential barrier.

The remainder of this paper is organized as follows. In section \ref{potentials}, we present the Lambert-W potential barrier, the hyperbolic tangent potential barrier, and the step potential barrier. In Section \ref{coefficients}, we find the solution of the Klein-Gordon equation with each of the potential barriers, as well as the reflection $R$ and transmission $T$ coefficients for each case. In section \ref{superradiance}, we discuss our results; finally, in section \ref{conclusion}, our conclusions are presented. 

\section{Potentials}
\label{potentials}

\bigskip
\subsection{Case 1: the step potential barrier} 

The step potential barrier is given by

\begin{equation}
V_{SP}(x)=
\left\{\begin{array}{c}
0, \quad x<0,\\
V_0, \quad x\geq 0
\end{array}
\right.
\end{equation}
where $V_0$ is the height of the barrier.

\bigskip
\subsection{Case 2: the hyperbolic tangent potential} 

The hyperbolic tangent potential barrier is defined as 

\begin{equation}
\label{potential}
V_{HT}(x)=\dfrac{V_0}{2}\left[\tanh(b\,x)+1\right],
\end{equation}
where $V_0$ represents the height of the potential and $b$ gives the smoothness of the curve.

\bigskip
\subsection{Case 3: the Lambert-W potential barrier} 

The asymmetric Lambert-W potential barrier is defined as

\begin{equation}
V_{LW}(x)=\dfrac{V_0}{1+\textnormal{W}\left(e^{-x/\sigma}\right)},
\end{equation}
where $V_0$ is the height of the barrier and $\sigma$ gives the smoothness of the potential barrier.

\begin{figure}[htbp]
\begin{center}
\includegraphics[scale=0.40]{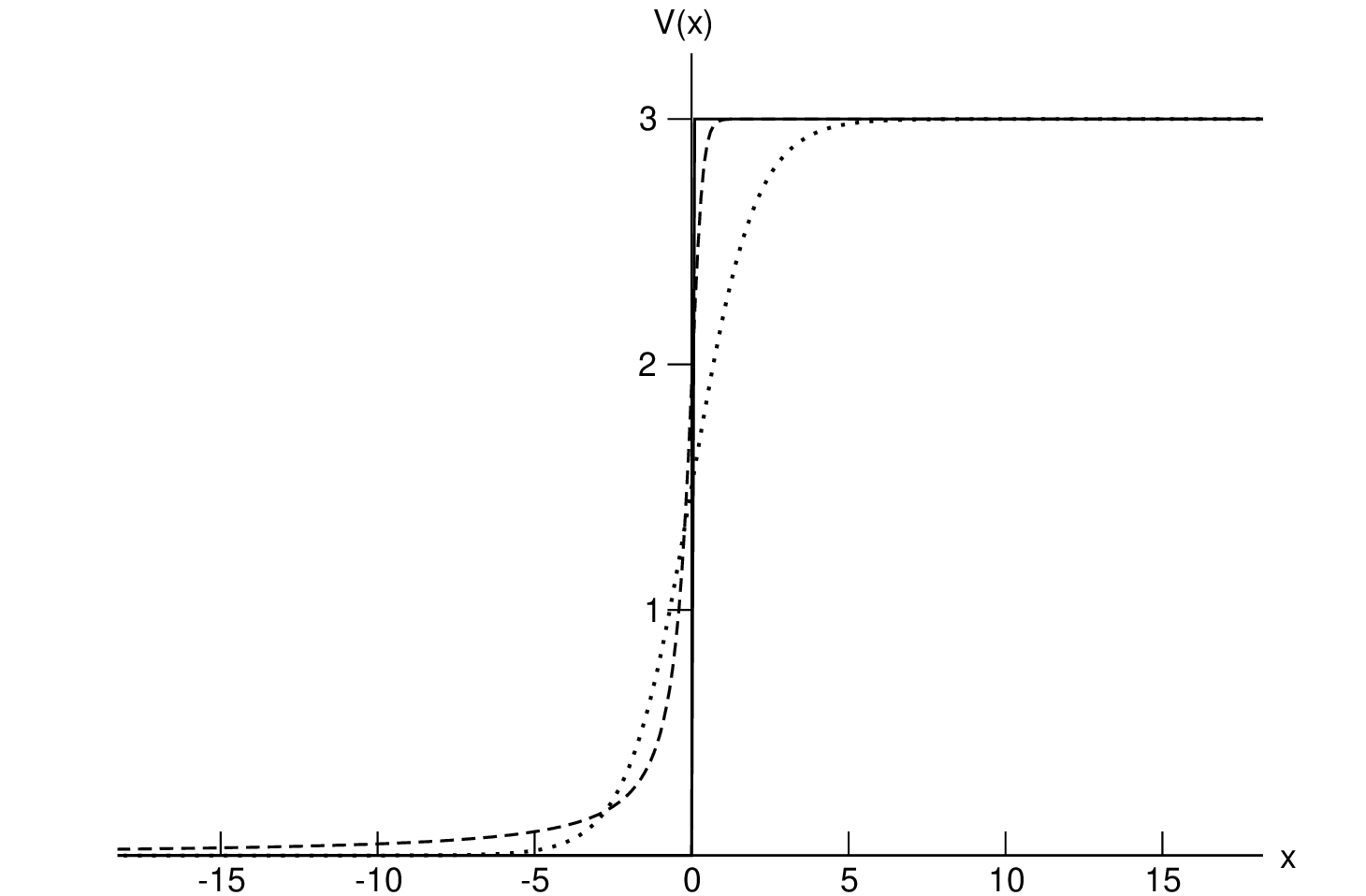}
\caption{\label{fig_pot}{Comparison between potential barriers; in all cases, $V_0=3$. Lambert-W potential barrier with $\sigma=0.15$ (dotted line), hyperbolic tangent potential barrier with $b=0.5$ (dashed line), and step potential barrier with $V_0=3$ (solid line)}}.
\end{center}
\end{figure}

\bigskip
The forms of these potentials barrier are shown in Fig. (\ref{fig_pot}). We can observe that as $b \rightarrow \infty$, the hyperbolic tangent potential reduces to the step potential. Because it is an asymmetric potential, the Lambert-W potential barrier does not reduce to the step potential for $\sigma \ll 1$.

\vspace{-0.7cm}
\section{Reflection and Transmission coefficients}
\label{coefficients}

\bigskip
\subsection{Case 1: the step potential barrier}

The reflection and transmission coefficients for the step potential barrier are given by \cite{wachter:2011}

\begin{equation}
\label{RSP}
R_{SP}=\left|\dfrac{(\mu-\nu)}{(\mu+\nu)}\right|^2,
\end{equation}

\begin{equation}
\label{TSP}
T_{SP}=\dfrac{\mu}{\nu}\left|\dfrac{2\nu}{\mu+\nu}\right|^2,
\end{equation}

\noindent
where $\nu=\sqrt{E^2-m^2}$ and $\mu=\sqrt{(E-V_0)^2-m^2}$.

\begin{figure}[!th]
\begin{center}
\subfigure[]{\includegraphics[scale=0.24]{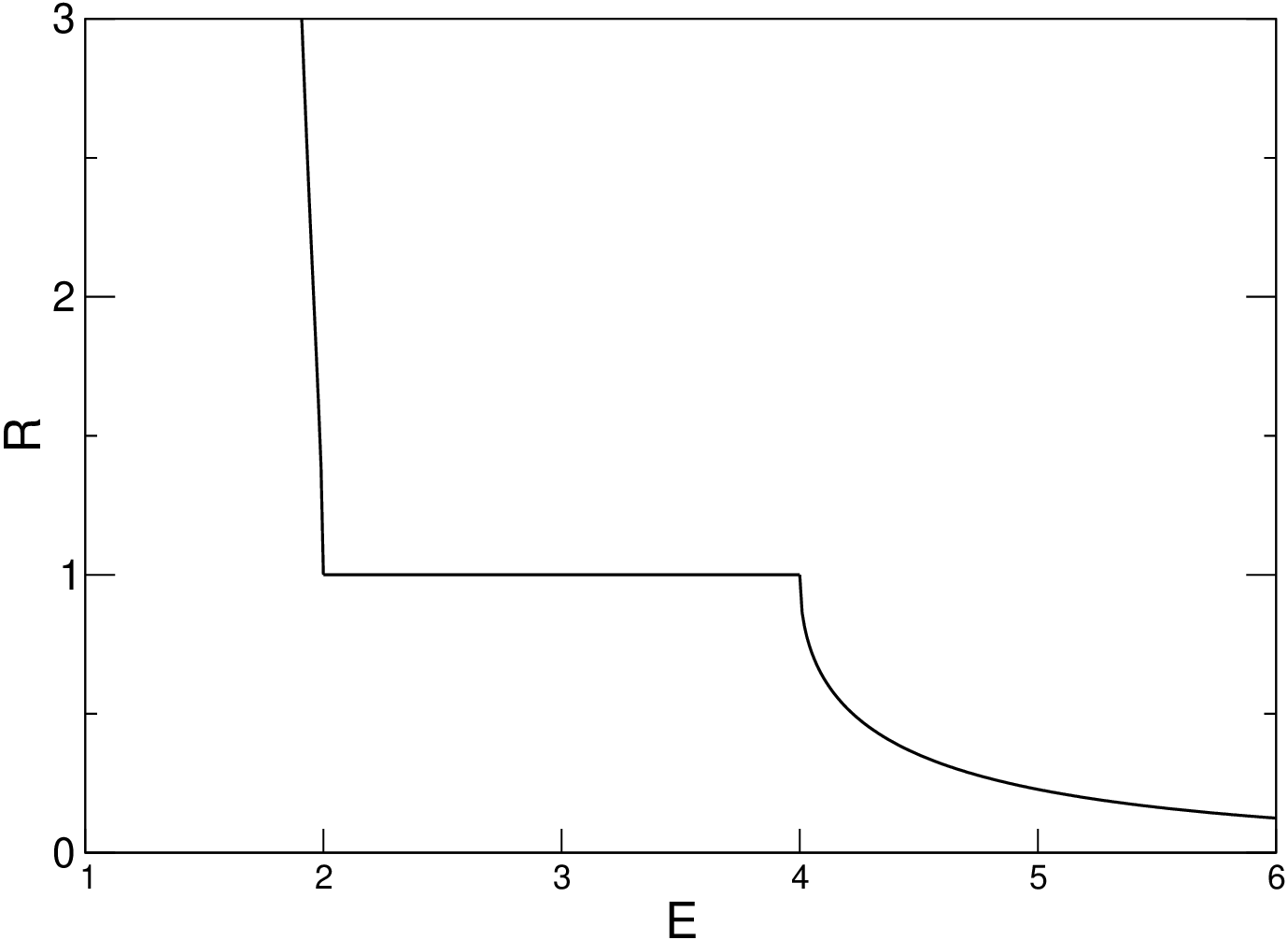}}
\hspace{1cm}\subfigure[]{\includegraphics[scale=0.24]{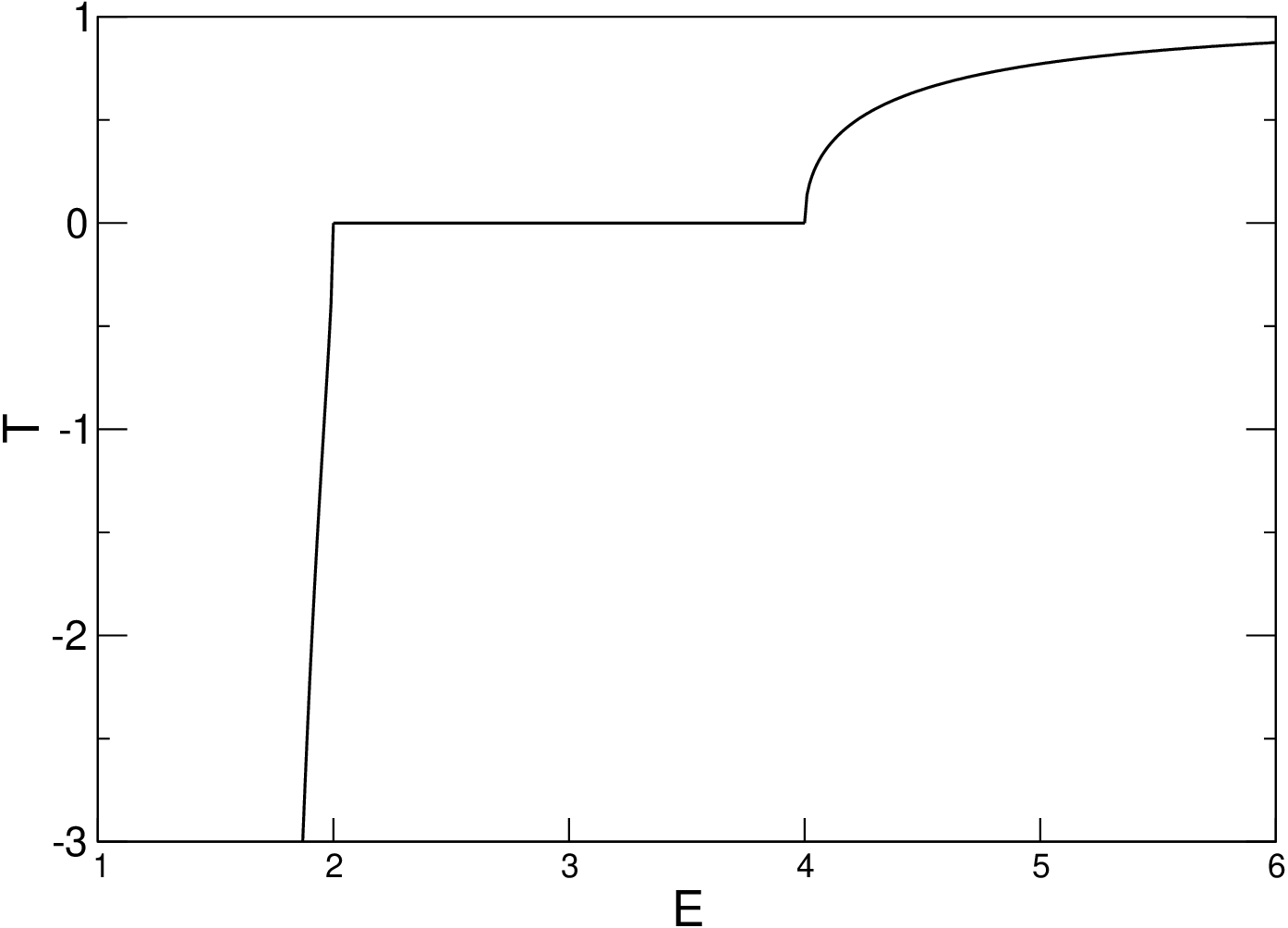}}
\caption{\label{TRSP} Reflection $R$ and transmission $T$ coefficients with varying energy $E$ for the relativistic step potential barrier with $V_0=3$ and $m=1$.}
\end{center}
\label{SP}
\end{figure}

\bigskip
The behavior of the reflection and transmission coefficients for the step potential barrier is represented in Fig. \ref{TRSP}(a) and Fig. \ref{TRSP}(b).

\bigskip
\subsection{Case 2: the hyperbolic tangent potential barrier} 

To consider the scattering solutions, we solve the differential equation  

\begin{equation}
\label{eq_x_1}
\frac{d^2\phi(x)}{dx^2}+\left\{\left[E-\dfrac{V_0}{2}\left[\tanh(bx)+1\right]\right]^2-m^2\right\}\phi(x)=0.
\end{equation}

Writing the hyperbolic tangent potential barrier in terms of the exponential functions, we obtain

\begin{equation}
\label{eq_x_2}
\frac{d^2\phi(x)}{dx^2}+\left[\left(E-\dfrac{V_0\, e^{2bx}}{1+e^{2bx}}\right)^2-m^2\right]\phi(x)=0.
\end{equation}

\medskip
By substituting $y=-e^{2bx}$ , Eq. (\ref{eq_x}) becomes

\begin{equation}
\label{eq_y)}
4b^2 y \frac{d}{dy}\left[y\frac{d\phi(y)}{dy}\right]+\left[\left(E+\frac{V_0y}{1-y} \right)^2-m^2 \right]\phi(y)=0.
\end{equation}

\medskip
Substituting $\phi_(y)=y^\alpha(1-y)^\beta f(y)$, Eq. (\ref{eq_y)}) reduces to the hypergeometric differential equation

\begin{equation}
\label{eq_hyper}
y(1-y)f''+[(1+2\alpha)-(2\alpha+2\beta+1)y]f'-(\alpha+\beta-\gamma)(\alpha+\beta+\gamma)f=0,
\end{equation}

where the primes denote the derivative with respect to $y$; the parameters $\alpha$, $\beta$, and $\gamma$ are

\begin{eqnarray}
\label{alpha}
\alpha&=&i\nu \,\,\, \textnormal{with} \,\,\, \nu=\frac{\sqrt{E^2-m^2}}{2b},\\
\label{beta}
\beta&=&\lambda \,\,\, \textnormal{with} \,\,\, \lambda=\frac{b+\sqrt{b^2-V_0^2}}{2b},\\
\label{gamma}
\gamma&=&i\mu \,\,\, \textnormal{with} \,\,\, \mu=\frac{\sqrt{(E-V_0)^2-m^2}}{2b}.
\end{eqnarray}

\medskip

Eq. (\ref{eq_hyper}) has a general solution in terms of Gauss hypergeometric functions $_2F_1(\mu,\nu,\lambda;y)$ \cite{rojas:2014a}

\begin{eqnarray}
\label{sol_y}
\nonumber
\phi(y)&=&C_1 y^\alpha \left(1-y\right)^\beta\, _2F_1\left(\alpha+\beta-\gamma,\alpha+\beta+\gamma,1+2\alpha;y\right)\\
&+&C_2 y^{-\alpha} \left(1-y\right)^\beta\, _2F_1\left(-\alpha+\beta-\gamma,-\alpha+\beta+\gamma,1-2\alpha;y\right).
\end{eqnarray}

\medskip
In terms of the variable $x$, Eq. (\ref{sol_y}) becomes

\begin{eqnarray}
\label{sol_x}
\nonumber
\phi(x)&=&c_1 \left(-e^{2bx}\right)^{i\nu} \left(1+e^{2bx}\right)^\lambda\, _2F_1\left(i\nu+\lambda-i\mu,i\nu+\lambda+i\mu,1+2 i\nu;-e^{2bx}\right)\\
\nonumber
&+&c_2 \left(-e^{2bx}\right)^{-i\nu} \left(1+e^{2bx}\right)^\lambda\, _2F_1\left(-i\nu+\lambda+i\mu,-i\nu+\lambda-i\mu,1-2 i\nu;-e^{2bx}\right).\\
\end{eqnarray}

\medskip
From Eq. (\ref{sol_x}), the incident and reflected waves are

\begin{equation}
\label{phi_inc}
\phi_{\textnormal{inc}}(y)=d_1 \left(1+e^{2bx}\right)^\lambda e^{2ib\nu x}\, _2F_1\left(i\nu+\lambda-i\mu,i\nu+\lambda+i\mu,1+2 i\nu;-e^{2bx}\right).
\end{equation}

\begin{equation}
\label{phi_ref}
\phi_{\textnormal{ref}}(y)=d_2 \left(1+e^{2bx}\right)^\lambda e^{-2ib\nu x}\, _2F_1\left(-i\nu+\lambda+i\mu,-i\nu+\lambda-i\mu,1-2 i\nu;-e^{2bx}\right).
\end{equation}

\medskip
Using the relationship \cite{abramowitz:1965}

\begin{eqnarray}
\label{relation}
\nonumber
_2F_1(a,b,c;z)&=&\frac{\Gamma(c)\Gamma(b-a)}{\Gamma(b)\Gamma(c-a)}(-z)^{(-a)}\,_2F_1(a,1-c+a,1-b+a;z^{-1})\\
\nonumber
&+&\frac{\Gamma(c)\Gamma(a-b)}{\Gamma(a)\Gamma(c-b)}(-z)^{(-b)}\,_2F_1(b,1-c+b,1-a+b;z^{-1}).\\
\end{eqnarray}

\medskip
The transmitted wave becomes

\begin{equation}
\label{phi_trans}
\phi_{\textnormal{trans}}(x)= d_3 e^{-2b\lambda x} \left(1+e^{2bx}\right)^\lambda e^{2ib\mu x}\, _2F_1\left(i\nu+\lambda-i\mu,-i\nu+\lambda-i\mu,1-2 i\mu;-e^{-2bx}\right).
\end{equation}

\medskip
As the incident wave is equal to the sum of the transmitted wave and the reflected wave,

\begin{equation}
\label{sumatoria}
\phi_{\textnormal{inc}}(x)=A\,\phi_{\textnormal{trans}}(x)+B\,\phi_{\textnormal{ref}}(x).
\end{equation}

\medskip
We use the relationship (\ref{relation}) and the equation for $\phi_{\textnormal{trans}}(x)$ to find

\begin{equation}
\label{phi_inc_def}
\phi_{\textnormal{inc}}(x)= A \left(1+e^{2bx}\right)^\lambda e^{2ib\nu x}\, _2F_1\left(i\nu+\lambda-i\mu,i\nu+\lambda+i\mu,1+2 i\nu;-e^{2bx}\right).
\end{equation}

\begin{equation}
\label{phi_ref_def}
\phi_{\textnormal{ref}}(x)=B \left(1+e^{2bx}\right)^\lambda e^{-2ib\nu x}\, _2F_1\left(-i\nu+\lambda+i\mu,-i\nu+\lambda-i\mu,1-2 i\nu;-e^{2bx}\right).
\end{equation}

\medskip
where the coefficients $A$ and $B$ in Eqs. (\ref{phi_inc_def}) and (\ref{phi_ref_def}) can be expressed in terms of the Gamma function as

\begin{equation}
\label{A}
A=\frac{\Gamma(1-2i\mu)\Gamma(-2i\nu)}{\Gamma(-i\nu+\lambda-i\mu)\Gamma(1-i\nu-\lambda-i\mu)}.
\end{equation}

\begin{equation}
\label{B}
B=\frac{\Gamma(1-2i\mu)\Gamma(2i\nu)}{\Gamma(i\nu+\lambda-i\mu)\Gamma(1+i\nu-\lambda-i\mu)}.
\end{equation}

\medskip
As $x\rightarrow \pm \infty$, $V \rightarrow \pm a$ and the asymptotic behavior of Eqs. (\ref{phi_trans}), (\ref{phi_inc_def}), and  (\ref{phi_ref_def}) are plane waves with the relationship of dispersion $\nu$ and $\mu$,

\begin{eqnarray}
\label{phi_asym}
\phi_{\textnormal{inc}}(x)&=& A e^{2ib\nu x},\\
\phi_{\textnormal{ref}}(x)&=& B e^{-2ib\nu x},\\
\phi_{\textnormal{trans}}(x)&=& e^{2ib\mu x}.
\end{eqnarray}

\medskip
To find $R$ and $T$, we used the definition of the electrical current density for the one-dimensional Klein-Gordon equation \cite{greiner:1987}

\begin{equation}
\label{current}
\vec{j}=\frac{i}{2}\left(\phi^*\vec{\nabla}\phi-\phi\vec{\nabla}\phi^*\right)
\end{equation}

\medskip
The current as $x \rightarrow -\infty$ can be decomposed to $j_\textnormal{L}=j_\textnormal{inc}-j_\textnormal{refl}$, where $j_\textnormal{inc}$ is the incident current and $j_\textnormal{ref}$ is the reflected one. Analogously, on the right side, as $x \rightarrow \infty$, the current is $j_\textnormal{R}=j_\textnormal{trans}$, where $j_\textnormal{trans}$ is the transmitted current \cite{rojas:2005}.

The reflection coefficient $R$, and the transmission coefficient $T$, in terms of the incident $j_\textnormal{inc}$, reflected $j_\textnormal{ref}$, and transmission $j_\textnormal{trans}$ currents are

\begin{equation}
\label{R}
R=\frac{j_\textnormal{ref}}{j_\textnormal{inc}}=\frac{|B|^2}{|A|^2}.
\end{equation}

\begin{equation}
\label{T}
T=\frac{j_\textnormal{trans}}{j_\textnormal{inc}}=\frac{\mu}{\nu}\frac{1}{|A|^2}.
\end{equation}

\begin{figure}[!th]
\begin{center}
\subfigure[]{\includegraphics[scale=0.24]{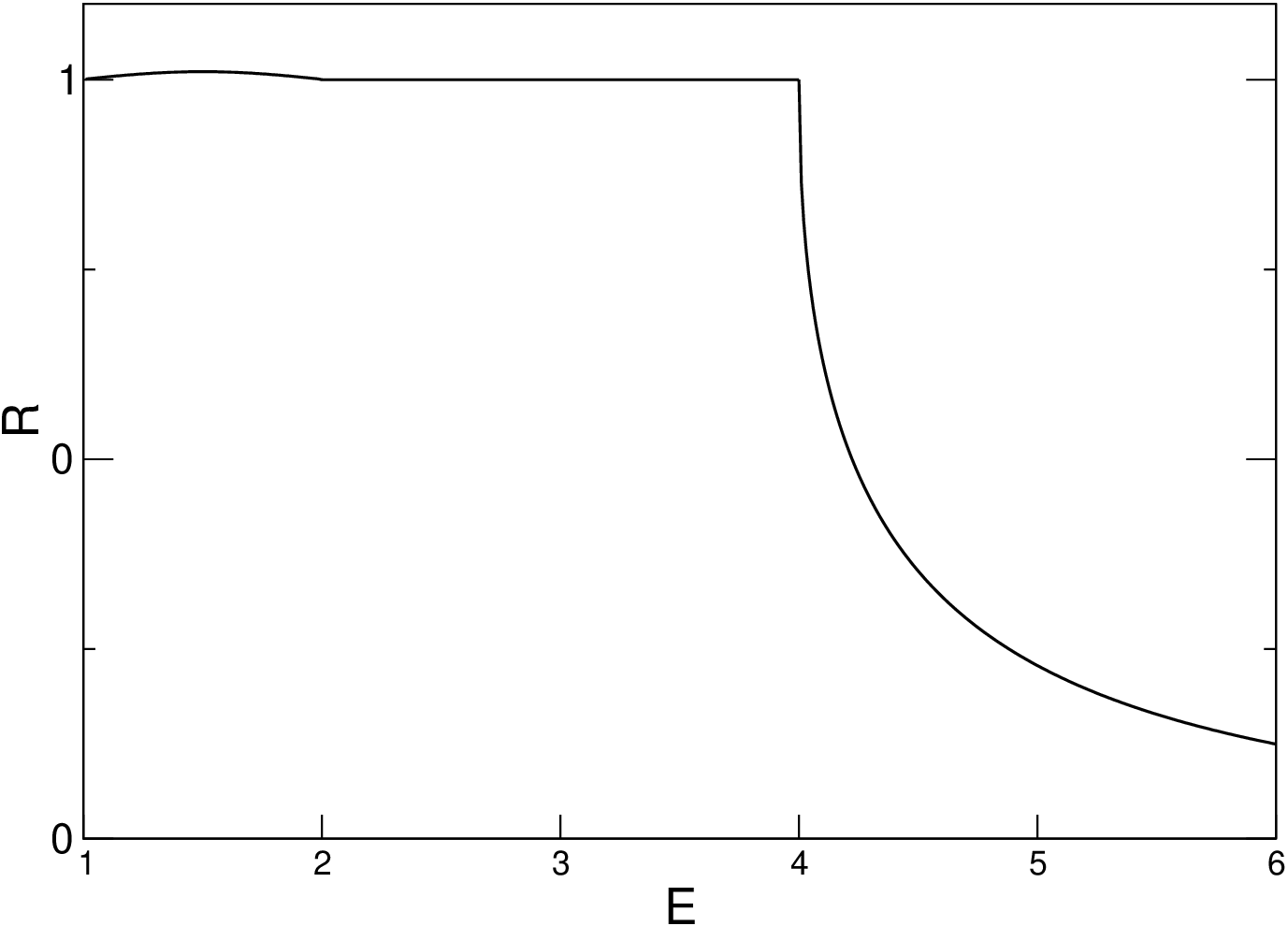}}
\hspace{1cm}\subfigure[]{\includegraphics[scale=0.24]{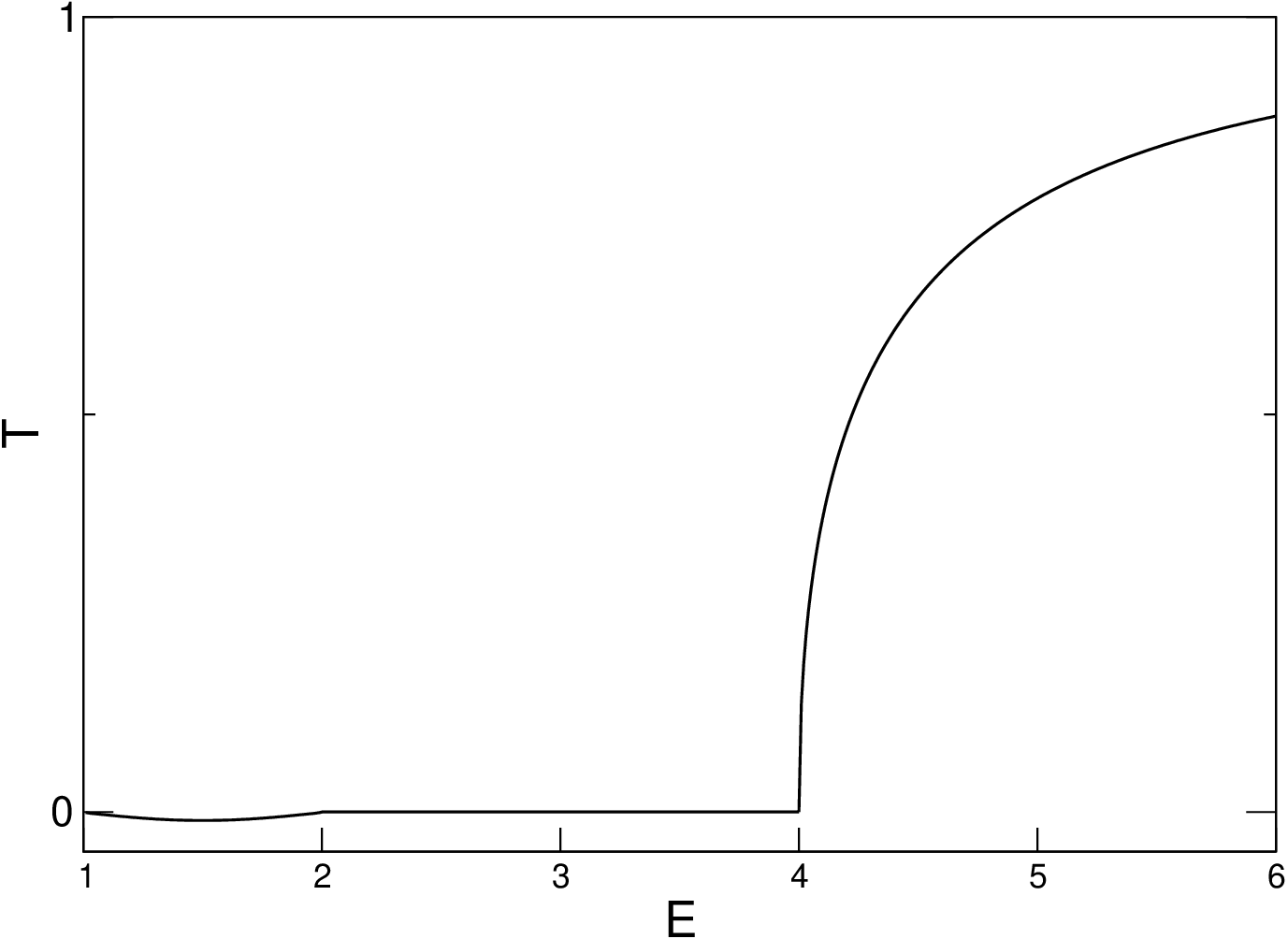}}
\caption{\label{TRHT} Reflection $R$ and transmission $T$ coefficients with varying energy $E$ for the relativistic hyperbolic tangent potential barrier with $V_0=3$, $b=0.5$, and $m=1$.}
\end{center}
\end{figure}

\bigskip
The behavior of reflection and transmission coefficients for the hyperbolic tangent potential barrier is represented in Fig. \ref{TRHT}(a) and Fig. \ref{TRHT}(b).

\bigskip

\subsection{Case 3: the Lambert-W potential barrier}

The one-dimensional Klein-Gordon equation to be solved is \cite{greiner:1987,wachter:2011}

\begin{center} 
\begin{equation}
\label{klein}
\frac{\mathrm{d}^2\phi(x)}{\mathrm{d}x^2}+\left\{\left[E-V(x) \right]^2-m^2 \right\}\phi(x)=0,
\end{equation}
\end{center}
where $E$ is the energy, $V(x)$ is the potential, and $m$ is the mass of the particle.

\bigskip
To consider the scattering solutions, we solve the differential equation  

\begin{equation}
\label{eq_x}
\frac{\mathrm{d}^2\phi(x)}{\mathrm{d}x^2}+\left\{\left[E-\dfrac{V_0}{1+\textnormal{W}\left(e^{-x/\sigma} \right)}\right]^2-m^2\right\}\phi(x)=0.
\end{equation}

\bigskip
On making the substitution $y=-\textnormal{W}\left(e^{-x/\sigma}\right)$, Eq. \eqref{eq_x} becomes

\begin{equation}
\label{eq_y}
\frac{\mathrm{d}^2\phi(y)}{\mathrm{d}y^2}+\dfrac{1}{y(1-y)}\dfrac{\mathrm{d}\phi}{\mathrm{d}y}+\sigma^2\left\{\dfrac{\left[E(1-y)-V_0\right]^2-m^2\left(1-y\right)^2}{y^2}\right\}=0.
\end{equation}

\bigskip
We propose the following solution to Eq. \eqref{eq_y}:

\begin{equation}
\label{phi_y}
\phi(y)=e^{\sfrac{\alpha}{2} y} y^{\sfrac{\beta}{2}} f(y),
\end{equation}
where $f(y)$ is the solution of the confluent Heun equation \cite{ronveaux:1995},

\bigskip\bigskip
Using the variable change Eq. \eqref{phi_y}, the differential equation \eqref{eq_y} becomes

\medskip
\begin{equation}
\label{eq_heunC}
\frac{\mathrm{d}^2f(y)}{\mathrm{d}y^2}-\dfrac{\left[-\alpha y^2+(-\beta+\alpha-\gamma-2)y+\beta+1\right]}{y(y-1)}\frac{\mathrm{d}f(y)}{\mathrm{d}y}-\dfrac{\left\{ \left[(-\beta-\gamma-2)\alpha-2\delta \right]y+(\beta+1)\alpha+(-\gamma-1)\beta-2\eta-\gamma\right\}}{2y(y-1)} f(y)=0,
\end{equation}
where $\alpha$, $\beta$,  $\gamma$, $\delta$, and $\eta$ are given by:

\begin{eqnarray}
\label{alpha}
\alpha&=& 2\sigma \sqrt{m^2-E^2},\\
\beta&=& =2\sigma\sqrt{m^2-E^2+2EV_0-V_0^2},\\
\gamma&=&-2,\\
\delta&=&2\sigma^2(m^2-E^2+EV_0),\\
\eta&=& 1-2\sigma^2(m^2-E^2+EV_0).
\end{eqnarray}

\medskip
Eq. (\ref{eq_heunC}) has a general solution in terms of the confluent Heun functions \cite{ronveaux:1995}

\begin{eqnarray}
\label{f_y}
f(y)&=&c_1 \textnormal{HeunC}(\alpha,\beta,\gamma,\delta,\eta,y)+ c_2 y^{-\beta} \textnormal{HeunC}(\alpha,-\beta,\gamma,\delta,\eta,y),
\end{eqnarray}
then Eq. \eqref{phi_y} becomes,

\begin{eqnarray}
\label{phi_y}
\phi(y)&=&c_1 e^{\sfrac{\alpha}{2} y} y^{\sfrac{\beta}{2}}\textnormal{HeunC}(\alpha,\beta,\gamma,\delta,\eta,y)+ c_2 e^{\sfrac{\alpha}{2} y} y^{-\sfrac{\beta}{2}} \textnormal{HeunC}(\alpha,-\beta,\gamma,\delta,\eta,y).
\end{eqnarray}

\bigskip
Finally, in terms of the variable $x$, the solution of Eq. \eqref{eq_x} becomes

\begin{eqnarray}
\label{phi_x}
\nonumber
\phi(x)&=&c_1 e^{-\frac{\alpha}{2} \textnormal{W}\left(e^{-x/\sigma}\right)} \textnormal{W}\left(e^{-x/\sigma}\right)^{\sfrac{\beta}{2}}\textnormal{HeunC}\left[\alpha,\beta,\gamma,\delta,\eta,-\textnormal{W}\left(e^{-x/\sigma}\right)\right]\\
\nonumber
&+&c_2 e^{-\frac{\alpha}{2} \textnormal{W}\left(e^{-x/\sigma}\right)} \textnormal{W}\left(e^{-x/\sigma}\right)^{-\sfrac{\beta}{2}}\textnormal{HeunC}\left[\alpha,-\beta,\gamma,\delta,\eta,-\textnormal{W}\left(e^{-x/\sigma}\right)\right] .\\
\end{eqnarray}

For the Lambert-W potential barrier, A. M. Ishkhanyan has found the following reflection coefficient \cite{ishkhanyan:2016d},

\begin{equation}
\label{RLW}
R_{LW}=e^{-2\pi\sigma \mu}\dfrac{\sinh\left[\frac{\pi\sigma}{2\nu}\left(\nu-\mu\right)^2\right]}{\sinh\left[\frac{\pi\sigma}{2\nu}\left(\nu+\mu\right)^2\right]},
\end{equation}
where for a relativistic particle, $\nu=\sqrt{E^2-m^2}$ and $\mu=\sqrt{(E-V_0)^2-m^2}$. 

\begin{figure}[!th]
\begin{center}
\subfigure[]{\includegraphics[scale=0.24]{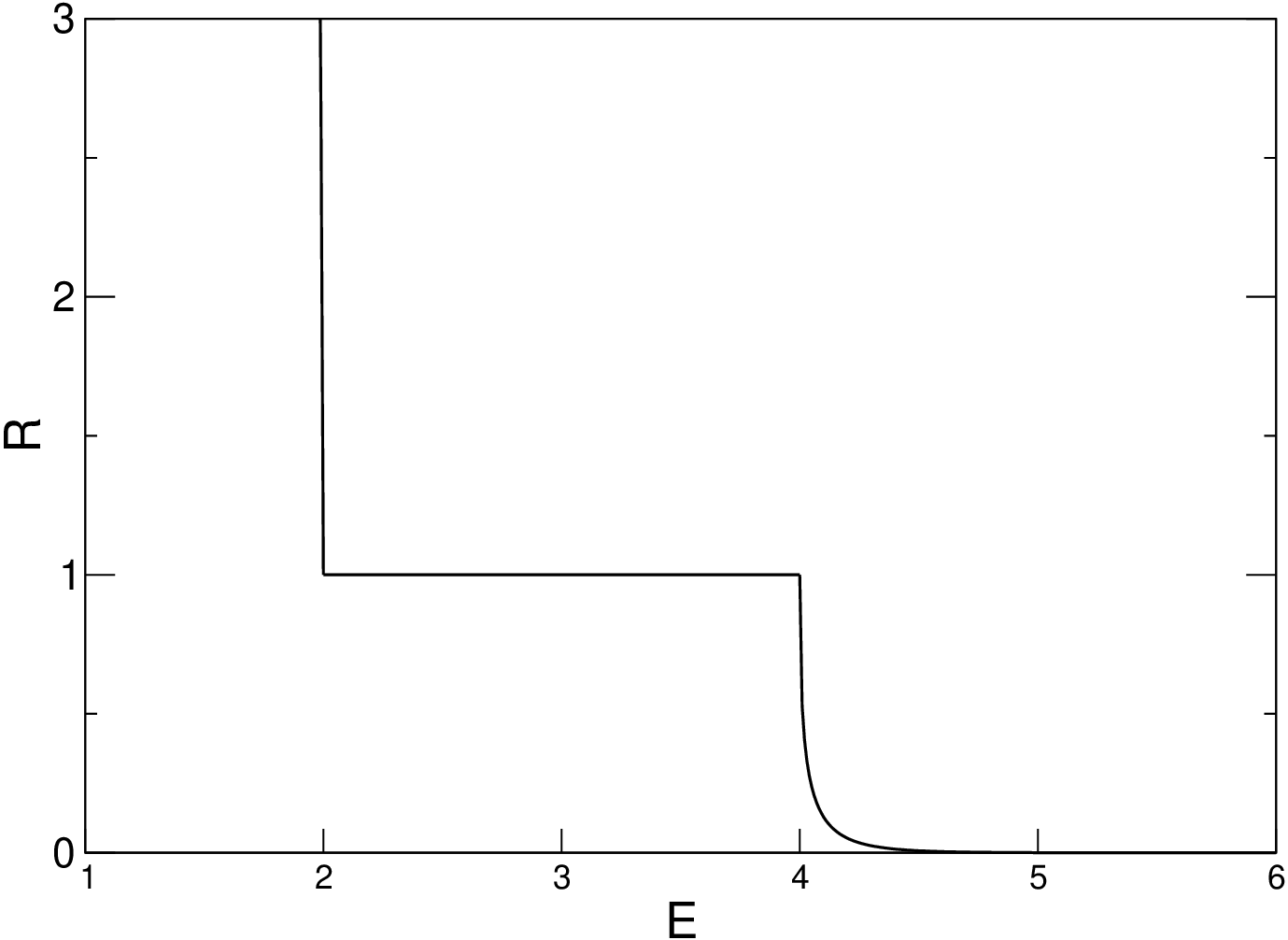}}
\hspace{1cm}\subfigure[]{\includegraphics[scale=0.24]{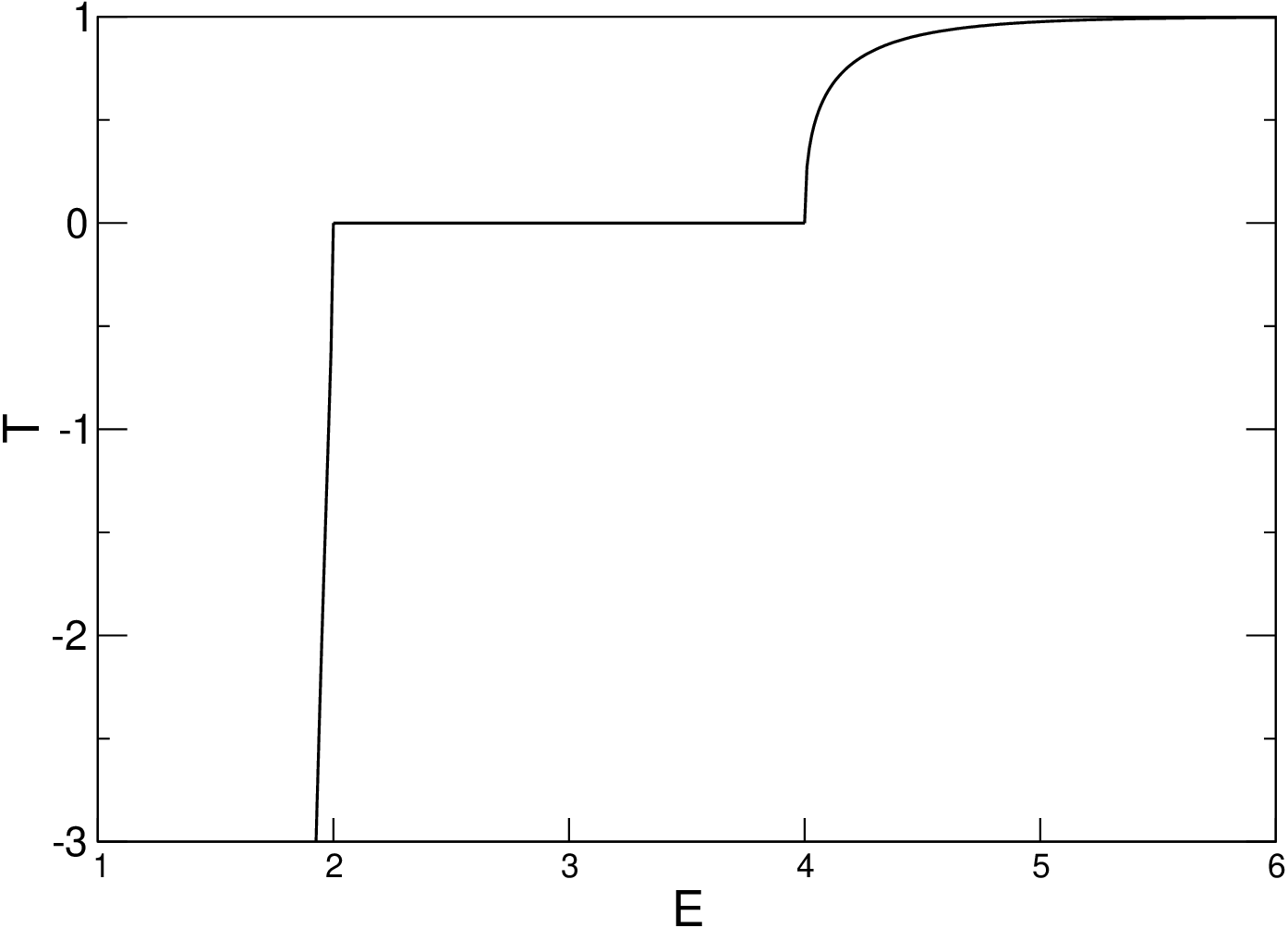}}
\caption{\label{TRLW} Reflection $R$ and transmission $T$ coefficients with varying energy $E$ for the relativistic Lambert-W potential barrier with $V_0=3$, $\sigma=0.15$, and $m=1$.}
\end{center}
\end{figure}

\bigskip
The behavior of the reflection and transmission coefficients for the Lambert-W potential barrier is represented in Fig. \ref{TRLW}(a) and Fig. \ref{TRLW}(b).

\section{Results}
\label{superradiance}

The dispersion relations $\nu'^2=E^2-m^2$ and $\mu'^2=(E+V_0)^2-m^2$ do not determine the sign of $\nu'$ and $\mu'$. These relations must be positive because they correspond to an incident particle moving from left to right, and their sign depends on the group velocity, 
which is calculated by taking the derivative of each dispersion relation with respect to the energy $E$ \cite{calogeracos:1999a}

\begin{equation}
 \label{group_nu}
 \frac{dE}{d\nu'}=\frac{\nu'}{E}\geq 0,
\end{equation}

\begin{equation}
 \label{group_muu}
 \frac{dE}{d\mu'}=\frac{\mu'}{E-V_0}\geq 0.
\end{equation}

\medskip
For these potentials, we have three different regions:

\subsection{Region 1: $V_0-m>E>m$} 

\bigskip
In the region $\mu'<0$ and $\nu'>0$, both terms are real and the quotient $\mu'/\nu'<0$, which implies that the transmitted wave oscillates in region 1. Because $\mu'<0$, in this region, $T\leq0$ and $R\geq 1$, the reflected current is greater than the incident current; therefore, superradiance occurs. The unitary relation $T+R=1$ is satisfied.

\subsection{Region 2: $V_0+m>E>V_0-m$} 

\bigskip
In this region, the dispersion relations $\mu$ and $\nu$ are purely imaginary, and the transmitted wave is exponentially damped down; therefore, $T=0$ and $R=1$; thus, $T+R=1$.

\subsection{Region 3: $E>V_0+m$} 

\bigskip
In this region, $\mu'>0$ and $\nu'>0$, both terms are real and the quotient $\mu'/\nu'>0$, which implies that the transmitted wave oscillates in region 3. Because $\mu'>0$, and $R>0$ and $T>0$, both are positive and satisfy the unitary relation $R+T=1$.

\bigskip
Figs. \ref{TRSP}(a) - \ref{TRLW}(b) show the reflection $R$ and transmission $T$ coefficients with various parameters for each potential. 
It is clear in the figures that, in the region $V_0-m>E>m$, the reflection coefficient $R$ is greater than one, whereas the transmission coefficient $T$ becomes negative; therefore, superradiance is observed \cite{cheng:2009,wagner:2010,molgado:2018}. In the region $V_0+m>E>V_0-m$, the reflection coefficient becomes equal to 1 and the transmission coefficient is equal to 0. In the region $E>V_0+m$, we observe that $1 \geq R \geq 0$ and $1 \geq T \geq 0$. In all regions, the coefficients $R$ and $T$ satisfy the unitary condition $T+R=1$.

\section{Conclusion}
\label{conclusion}

In this study, we investigated the phenomenon of superradiance in the Lambert-W potential barrier. We compared its solution with those of the hyperbolic tangent potential barrier and the step potential barrier. We compared the reflection $R$ and transmission $T$ coefficients for these three different potentials, showing that for the region where $V_0-m>E>m$, the phenomenon of superradiance occurs.

\section{Acknowledgment}

We want to express our gratitude to Dr. A. M. Ishkhanyan for useful discussions.


\end{document}